\input harvmac.tex 

\def\Asym#1#2{\vcenter{\vbox{\drawbox{#1}{#2}
              \kern-#2pt       
              \drawbox{#1}{#2}}}}

\Title{\vbox{\rightline{hep-th/9802203} \rightline{CERN-TH/98-62}
\rightline{UCLA/98/TEP/6}
}}
{\vbox{\centerline{On N=8 Supergravity in $AdS_5$}
\centerline{and N=4 Superconformal Yang-Mills theory}}}


\centerline{Sergio Ferrara} \smallskip{\it
\centerline{CERN Geneva, Switzerland}}
\centerline{\tt
Sergio.Ferrara@cern.ch}

\vskip .3cm
\centerline{Christian Fr\o nsdal} \smallskip{\it
\centerline{Physics Department, University of California, Los Angeles, CA
90090-1547}}
\centerline{\tt
fronsdal@physics.ucla.edu}

\vskip .3cm
\centerline{Alberto Zaffaroni} \smallskip{\it
\centerline{CERN Geneva, Switzerland}}
\centerline{\tt
Alberto.Zaffaroni@cern.ch}

\vskip .1in


\noindent
We discuss the spectrum of states of IIB supergravity on $AdS_5\times S^5$ in
a manifest $SU(2,2/4)$ invariant setting. The boundary fields are described in terms of $N=4$ superconformal Yang-Mills theory and the proposed correspondence
between supergravity in $AdS_5$ and superconformal invariant singleton theory
at the boundary is formulated in a $N=4$ superfield covariant language.
\vskip 3truecm
\noindent
CERN-TH/98-62
\Date{February 98}

\lref\malda{ J. M. Maldacena, {\it The Large N Limit of Superconformal Field Theories and Supergravity},  hep-th/9705104.}
\lref\maldatwo{ N. Itzhaki, J. M. Maldacena, J. Sonnenschein and S. Yankielowicz, {\it Supergravity and The Large N Limit of Theories With Sixteen Supercharges}, hep-th/9802042.}
\lref\witten{E. Witten, {\it Anti-de Sitter Space And Holography}, hep-th/9802150.}
\lref\pol{S. S. Gubser, I. R. Klebanov and A. M. Polyakov, {\it Gauge Theory Correlators from Non-Critical String Theory}, hep-th/9802109.}
\lref\fer{S. Ferrara and  C. Fronsdal, {\it  Conformal Maxwell theory as a singleton field theory on $ADS_5$, IIB three branes
     and duality}, hep-th/971223.}
\lref\fz{S. Ferrara and B. Zumino, Nucl. Phys. B87 (1975) 207.}
\lref\fertwo{S. Ferrara and  C. Fronsdal, {\it Gauge Fields as Composite Boundary Excitations}, hep-th/9802126.}
\lref\van{H. J. Kim, L. J. Romans and P. van Nieuwenhuizen, {\it The Mass Spectrum Of Chiral N=2 D=10 supergravity on $S^5$}, Phys. Rev. D23 (1981) 1278.}
\lref\roo{E. Bergshoeff, M. De Roo and B de Wit, {\it Extended Conformal Supergravity}, Nucl. Phys. B182 (1981) 173.}
\lref\stelle{P. Howe, K. S. Stelle and P. K. Townsend, {\it Supercurrents}, Nucl. Phys. B192 (1981) 332.}
\lref\kleb{C. W. Gibbons and P. K. Townsend, Phys. Rev. Lett. 71 (1993) 3754; M. P. Blencowe and M. J. Duff, Phis. Lett. B203 (1988) 229; Nucl. Phys B310 (1988), 389; M. J. Duff, Class. Quantum Grav. 5 (1988) 189; E. Bergshoeff, M. J. Duff, C. N. Pope and E. Sezgin, Phys. Lett. B199 (1988) 69; H. Nicolai, E. Sezgin and Y. Tanii, Nucl. Phys B305 (1988) 483.}
\lref\sken{K. Sfetsos and K. Skenderis, {\it Microscopic derivation of the Bekenstein-Hawking entropy formula for
     non-extremal black holes}, hep-th/9711138.}
\lref\skentwo{H. J. Boonstra, B. Peeters and K. Skenderis, {\it Branes and anti-de Sitter spacetimes}, hep-th/9801076.}
\lref\kall{P. Claus, R. Kallosh and A. Van Proeyen, {\it M 5-brane and superconformal (0,2) tensor multiplet in 6 dimensions}, hep-th/9711161.}
\lref\kalltwo{ R. Kallosh, J. Kumar and  A. Rajaraman, {\it Special Conformal Symmetry of Worldvolume Actions}, hep-th/9712073.}
\lref\kallthree{P. Claus, R. Kallosh, J. Kumar, P. Townsend and A. Van Proeyen, {\it Conformal theory of M2, D3, M5 and D1+D5 branes}, hep-th/9801206.}
\lref\ooguri{G. T. Horowitz and H. Ooguri, {\it Spectrum of Large N Gauge Theory from Supergravity}, hep-th/9802116.}
\lref\gun{M. Gunaydin and D. Minic {\it Singletons, Doubletons and M-theory}, hep-th/9802047}
\lref\silv{S. Kachru and  E. Silverstein, {\it 4d Conformal Field Theories and Strings on Orbifolds}, hep-th/9802183.}
\lref\fztwo{S. Ferrara and B. Zumino, Nucl. Phys. B134 (1978) 301.}
\lref\guntwo{M. Gunaydin, L. J. Romans and N. P. Warner, Phys. Lett. 154B (1985) 268; M. Pernici, K. Pilch and P. van Nieuwenhuizen, Nucl. Phys. B259 (1985) 460.}
\lref\fr{M. Flato and C. Fronsdal, J. Math. Phys. 22 (1981) 1100; Phys. Lett. B172 (1986) 412.}
\lref\frtwo{M. Flato and C. Fronsdal, Lett. Math. Phys 2 (1978) 421; Phys. Lett. 97B (1980) 236.}
\lref\frthree{E. Angelopoulos, M. Flato, C. Fronsdal and D. Sternheimer, Phys. Rev. D23 (1981) 1278.}
\newsec{Introduction.}
In recent times, it has been conjectured that there is a close connection
between certain supergravity theories in $AdS_{p+2}$ and p-brane dynamics
on the $p+1$ world-volume\refs{\kleb,\malda,\sken,\skentwo,\kall,\kalltwo,\kallthree}. 

In certain limits, this relation has a plausible validity of application and it
implies, in a more algebraic setting, that certain field and string theories,
based on the same superalgebra, are {\it dual} to each other.
This was first conjectured in \malda, and further studied in \refs{\maldatwo,\ooguri,\gun}. In recent papers \refs{\pol,\witten}, a more precise definition for this duality was given, using a {\it functional approach} to Green function correlators, in analogy to the relation between string theory and target space dynamics.

In a recent set of papers, many properties of the possible interconnections started to be explored in three main directions: comparing theories of different
p-branes in different dimensions with the nearly horizon supergravity theory
in $AdS_{p+2}$ \maldatwo, computing correlators in the {\it boundary} theories
to get informations on the spectrum of the {\it bulk} theory \refs{\pol,\witten} or {\it viceversa}, using the group-theoretical relation of gauge fields in anti-de-Sitter space
with {\it singletons} in the boundary theories \refs{\fer,\fertwo}.

In particular, it was shown in \refs{\fer,\fertwo} that {\it massless gauge} fields in $AdS_n$ correspond to {\it composite boundary excitations} which, in a
field theory framework, correspond to the existence of some {\it conformal operators} in the boundary theory whose scaling dimensions do not get 
renormalized.

More recently, these expectations have been confirmed by dynamical calculations in $N=4$ superconformal $U(n)$ Yang-Mills theory for the case of D3 branes \refs{\pol,\witten} where horizon geometry is related to $N=8$ supergravity in $AdS_5$. Yang-Mills theory is the gauged $SU(4)-R$ symmetry of the four-dimensional boundary theory \refs{\fer,\fertwo,\guntwo}.

It is the aim of the present paper to expand further this setting by providing,
as promised in previous papers \fer, a manifest $N=4$ superconformal invariant 
formulation  of the {\it composite boundary excitations} as well as of the {\it field theory formulation} of the proposed duality, as recently suggested by
various authors.

The paper is organized as follows. In section 2, we give some basic facts on 
certain gauge fields on $AdS_5$, by showing that for these fields, quite independently of supersymmetry, there is a reasonable definition of {\it mass},
which coincides with what comes out from the supergravity literature. In section 3, we show the relevant {\it particle} multiplet of $SU(2,2/4)$ superalgebra
and the {\it field representations}. This in particular allows to check the spectrum of K-K excitations of type IIB on $AdS_5\times S^5$ in a manifest $N=4$
formalism. As an output, we show that fields with different masses but the same spin (including the elementary excitations), apparently coming from different towers of operators, are actually coming from the same $N=4$ supertower.
In section 4, we give an $N=4$ manifest formulation of the {\it functional approach} proposed in \refs{\pol,\witten}, to relate the bulk supergravity theory to
the boundary conformal field theory. The appendix summarizes some properties of conformal superfields.

\newsec{Unitary representations of particle states in $AdS_5$.}
Elementary particles in $AdS_{d+1}$ are classified by unitary, irreducible representations of $SO(d,2)$.
\lref\mack{G. Mack and A. Salam, Ann. Phys. 53 91969) 174.}
\lref\fe{S. Ferrara, Nucl. Phys B77 (1974) 73.}
\lref\fgg{S. Ferrara, A. F. Grillo and R. Gatto, Ann. Phys. 76 (1973) 161; Phys. Rev. D9, (1974) 3564.} 
To classify the states \refs{\fr,\frtwo,\frthree}, $SO(d,2)$ is decomposed with respect to its maximal compact subgroup $SO(d)\times SO(2)$. Representations with positive energy
are highest weight representations classified by the highest weights; that
is, once again by $SO(2) \times SO(d)$ quantum numbers,
the minimal energy $E_0$ and the spin. Thus, for d = 4, $SO(4)=SU(2)\times SU(2)$, and irreducible representations are classified by three labels, two spins
and the energy. We denote such representations by $D(E_0,J_1,J_2)$.

An important object, which will enter in the definition of a particle mass
in $AdS_5$ is the quadratic Casimir of $SO(4,2)$ which takes the value,
\eqn\casimir{E_0(E_0-4) + 2J_1(J_1+1) + 2J_2(J_2+1).}

Unitary representations lie within two unitary conformal bounds \refs{\mack,\fgg}: 1) $E_0\ge 2+ J_1 +J_2$
(if $J_1J_2\ne 0$) and 2) $E_0\ge 1+J$ (if $J_1J_2=0$). It was shown in
 \refs{\fer,\fertwo} that these bounds are precisely saturated by massless particles in anti-De Sitter (1) and massless particles at the boundary (2).

\lref\lao{M. Laoues, {\it Some properties of massless particles in arbitrary dimensions},
 Dijon (Univ. de Bourgogne, Lab. Gevrey) preprint, Jan 1998; E. Angelopoulos and M. Laoues, {\it Masslessness in n-dimensions},
Dijon (Univ. de Bourgogne, Lab. Gevrey) preprint, July 1997.}

Let us now consider gauge fields of interest in $AdS_5$; they are vectors, symmetric tensors and antisymmetric tensors. Imposing that the laplacian operator
substains a gauge symmetry for massless
tensor fields, we define the square {\it mass}
in terms of the value of the Casimir operator, shifted so as to be zero in
the massless case\foot{Investigation of masslessness in arbitrary dimensions has been recently the subject of careful analysis in \lao.}. The result is \fertwo,
\item{.} Vector field: $A_{\mu}$
\eqn\vec{D(E_0,{1\over 2},{1\over 2}):\qquad m^2=C_I=(E_0-1)(E_0-3)}
\item{.} Symmetric tensor: $g_{\mu\nu}=g_{\nu\mu}$
\eqn\tens{D(E_0,1,1):\qquad m^2=C_I-8=E_0(E_0-4)}
\item{.} Antisymmetric tensor: $A_{\mu\nu}=-A_{\nu\mu}$
\eqn\anti{D(E_0,1,0) \oplus D(E_0,0,1):\qquad m^2=C_I=(E_0-2)^2}

while for scalar fields, as usual,
\item{.} Scalars: $\phi$
\eqn\scal{D(E_0,0,0):\qquad m^2=C_I=E_0(E_0-4)}
and for fermions,
\item{.} Fermions of spin $3/2$: $\psi_\mu$
\eqn\grav{D(E_0,1,{1\over 2})+D(E_0,{1\over 2},1);\qquad m=E_0-2}
\item{.} Fermions of spin $1/2$: $\lambda$
\eqn\fermions{D(E_0,0,{1\over 2})+D(E_0,{1\over 2},0);\qquad m=E_0-2}
\lref\gunthree{M. Gunaydin and N. Marcus, Class. Quantum Grav. 2 (1985) L11.}

It is not surprising that all the K-K excitations, both massless and massive,
for type IIB supergravity on $AdS_5\times S^5$ \refs{\guntwo,\gunthree,\van}, verify
these formulae, for different values of $E_0$. Apparent different towers are due to the fact that supersymmetry relates particles with the same value of $(J_1,J_2)$ but different $E_0$ and $SU(4)$ assignment.

The unitarity bound $E_0=2+J_1+J_2$ corresponds to {\it conserved currents} in the boundary conformal field theory \fertwo. It corresponds  to {\it massless}
gauge fields in $AdS_5$; therefore we get that
\eqn\pesi{A_\mu:\qquad D(3,{1\over 2}, {1\over 2});\qquad\qquad g_{\mu\nu}:\qquad D(4,1,1);\qquad\qquad  A_{\mu\nu}: D(3,1,0)\oplus D(3,0,1).}
For scalars, since there is no gauge symmetry, there is no  unique definition of masslessness, but for the special value $m^2=0$ we have $E_0=4$; this
state will be associated to the (complex) dilaton, which, as expected, has the same $E_0$ of the graviton.

\lref\gunproc{ M. Gunaydin, Proc. Trieste conf. {\it Supermembranes and Physics in $2+1$
dimensions}, edited by M. J. Duff, C. N. Pope, E. Sezgin, World Scientific, 1990, 442; 
M. Gunaydin, B. E> W. Nilsson, G. Sierra and P. K. Townsend, Phys. Lett. B176 (1986) 45.}

The other exceptional representation is the singleton representation which saturates the other bound $E_0=1+J$. It essentially corresponds to a topological theory which lives at the boundary of $AdS_5$. For $J=1$, this gives the singleton Maxwell theory discussed in \fer. $J=0,1/2$ are the singletons discussed in
the Anti-de Sitter literature \refs{\fr,\frtwo,\frthree,\gun,\gunproc}; they corresponds to the irreducible representations $D(1,0,0),D(3/2,1/2,0)$ and $D(2,1,0)$  of $O(4,2)$. These fields are the building blocks of $N=4$ super
Yang-Mills theory on $M_4$. It is the D3 brane world-volume theory.

\newsec{Supermultiplets and superfields of $SU(2,2/4)$.}
In $N=4$ conformal supersymmetry, with fields defined at the boundary of
$AdS_5$, we have two basic multiplets, which are related to massless particles at the boundary (supersingleton) and massless particles in the bulk (supergraviton).

As proposed in \fer, following the conjecture of \malda, the supergraviton multiplet is a {\it composite boundary operator} of singleton fields at the boundary. It corresponds to gauge covariant bilinear composites, described by a supermultiplet of currents  at the boundary, as explained in \fertwo.

Let us give the $D(E_0,J_1,J_2)$ components of the multiplets:
\item {.} Supersingleton multiplet\foot{For singleton multiplets in $M_4$ antiparticle states are included.},
\eqn\singl{D(1,0,0|6)+D({3\over 2},{1\over 2},0|4)+D({3\over 2},0,{1\over 2}|\bar 4)+D(2,1,0|1)+D(2,0,1|1),}
where the fourth index in the D symbol is the $SU(4)$ representation. The supermultiplet contains the field content of an $N=4$ vector multiplet, a gauge field, four complex fermions in the {\bf 4} of $SU(4)$ and six scalars in the
{\bf 6} of $SU(4)$.

\lref\warner{M. Gunaydin, L. J. Romans and N. P. Warner, Nucl. Phys. B272 (1986) 598.}

\item{.} Supergraviton multiplet \foot{A list of the $AdS_5$ massless multiplets for various amount of supersymmetry can be found in \warner.},
\eqn\graviton{\eqalign{&D(4,1,1|1)+D({7\over 2},1, {1\over 2}|\bar 4)+ D({7\over 2},{1\over 2},1|4)+D(3, {1\over 2},{1\over 2}|15)+D(3,1,0|6_c)\cr +&D(3,0,1|\bar 6_c)
+D({7\over 2},{1\over 2},0|4)+D({7\over 2},0,{1\over 2}|\bar 4)+ D({5\over 2},{1\over 2},0|20)+ D({5\over 2},0,{1\over 2}|{\bar 20})\cr +&D(2,0,0|20_R)+D(3,0,0|10)
+D(3,0,0|{\bar 10})+D(4,0,0|1)+D(4,0,0|\bar 1),}}
which contains the graviton, eight gravitinos in the ${\bf 4}+ \bar {\bf 4}$
of $SU(4)$, fifteen vectors which gauge $SU(4)$, twelve antisymmetric tensors in the ${\bf 6}_c$, forty-eight spin $1/2$ fermions in the ${\bf 4}+\bar {\bf 4}+{\bf 20}+
\bar{\bf 20}$ and forty-two scalars in the ${\bf 20}_R+{\bf 10}+\bar {\bf 10}+{\bf 1}+\bar {\bf 1}$.

As anticipated, we see that there are spin $1/2$  and spin $0$ particles having different {\it masses}, since they have different $E_0$. However, the masses are completely fixed by $N=4$ supersymmetry.
To read the quantum number it sufficies to associate these fields to the $N=4$ conformal supercurrent multiplet $J_{[AB][CD]}$ (with $A=1,...,4$)\foot{$J_{[AB][CD]}$ stands for the expression $J_{\{[AB],[CD]\}}-{1\over 24}\epsilon_{ABCD}J_{[EF][GH]}\epsilon^{EFGH}$.}, whose complete component expansion was given in \roo, and whose $N=4$ superfield was given in \stelle. 

The multiplet starts with the traceless symmetric combination of $Tr \phi_{\{ l}\phi_{m\}}$ (with $i=1,...,6$). Since $\phi_l$ is a field with $E_0=1$, this operator has $E_0=2$.
As for the other scalars, those with $E_0=3$ in the ${\bf 10}$ of $SU(4)$ are  $\theta^2$ components of this superfield, while the singlet dilaton with $E_0=4$ is a $\theta^4$ component, as shown in \refs{\roo,\stelle}. In an analogous way, the spin $1/2$ with $E_0=5/2$ in the ${\bf 20}_c$ is a $\theta$ component while the spin $1/2$ in the ${\bf 4}$ with $E_0=7/2$ is a $\theta^3$ component. Finally, the
vector and antisymmetric tensors are in the $\theta^2$ component and the graviton in the $\theta^4$. The components with from five to eight $\theta$ are actually derivatives of lower components.

In \witten, some K-K towers of scalars in the spectrum of type IIB on $AdS_5\times S^5$ have
been identified with BPS composite fields in the $N=4$ Yang-Mills theory, as a check of the proposed duality of this theory with the world-volume theory of the D3 branes.   
It is not difficult at this point to 
perform the analysis of the spectrum in an $N=4$ manifestly covariant way and to identify also the composite fields associated with gauge fields, symmetric and antisymmetric tensors and fermions.
The K-K excitations of type IIB
on $AdS_5\times S^5$ are precisely given by taking $SU(n)$ gauge invariant polynomials
of the singleton supermultiplets, whose superfield expression is a superfield strenght $W_l=\Gamma_l^{[AB]}W_{[AB]}$, satisfying the constraints given in eq. (2.5) and (2.6) in \stelle.
The massless multiplet in $AdS_5$ corresponds to
\eqn\multi{J_{lm}=Tr W_{\{ l}W_{m\}},}
which is the {\it supercurrent} multiplet of \stelle, and it corresponds to the supergraviton multiplet we just discussed in detail. The singleton superfield $W_l$, which is Lie algebra valued in $SU(n)$, as well as the supercurrent superfield $J_{lm}$, satisfy some constraints, that can be found in \stelle.

If we read the towers of K-K excitations in \van, we see that, for a given $(J_1,J_2)$ they satisfy the main formulae given in eq. \vec, \tens, \anti, \scal, \grav\ and \fermions, but with different values of $E_0$ as given by the $N=4$ superfield
multiplication. 

Let us briefly check the massless multiplet and the first excitations. The conformal weight and the $SU(4)$ representation for the components
of the massless multiplet was given above. Using formula \scal, for the three
type of scalars with $E_0=2,3,4$ we expect to find in the supergraviton
multiplet scalars in the ${\bf 20}_R$ of $SU(4)$ with $m^2=-4$, scalars in the ${\bf 10}$ with $m^2=-3$ and singlet scalars with $m^2=0$. This is indeed the result obtained in \van, as can be easily checked by looking at table III in that reference and at the various figures where states in the supergraviton multiplet are
encircled. Using formulae \vec, \tens, since we have vectors and tensors with $E_0=3$, we reproduce the massless gauge fields
which fill the adjoint of $SU(4)$ and an antisymmetric tensor in the ${\bf 6}_c$
of $SU(4)$ with $m^2=1$ found in \van\ in the supergraviton multiplet. Using
eq. \grav, \fermions, also the masses for the fermions are easily shown to agree with
the ones found in \van; in particular it is clear that the two spin $1/2$ fermions in the ${\bf 20}_c$ and ${\bf 4}$ have different masses since they have different $E_0$. 

The massive K-K excitations can be obtained by considering higher degree polynomials of $W_l$. The $N=4$ superfield multiplication gives immediatly the dimension $E_0$ and the $SU(4)$ representation. It can be easily checked that the full tower of K-K states can be obtained in this way. We simply notice, as a check, that, in
complete agreement with the formulae which give the masses in terms of $E_0$, all the masses for
the scalar K-K modes in \van\ have the form $E_0(E_0-4)$ for some $E_0$, as in formula \scal, the masses for the
tensors have the form $(E_0-2)^2$ for some $E_0$, as in formula \tens, the masses for vectors have the form $(E_0-1)(E_0-3)$ for some $E_0$, as in formula
\vec, and so on. 
  
\newsec{On the CFT/AdS connection.}
In order to apply the $N=4$ formalism to the proposal of the $CFT/AdS$ equivalence, we recall more on the field representations of the $SU(2,2/4)$ algebra.
They are the Weyl superfield and the gravity gauge potential superfield. 

The Weyl superfield is a chiral conformal multiplet $W$ which satisfies a set of constraints  given in \stelle. Let us just remember that the analogous
multiplet in $N=1$ is $W_{\alpha\beta\gamma}$ and in $N=2$ is $W_{\alpha\beta}$ whose first components are the gravitino and graviphoton field strenght, respectively. In the $N=4$ case, the $W$ multiplet is a dimension 0 chiral superfield, whose first component is a complex scalar, the conformal dilaton $\phi_0$.

This superfield has a {\it prepotential} $V_{[AB][CD]}$ of conformal dimension $-6$ whose relation to the Weyl multiplet is \stelle,
\eqn\prep{ W=\bar D^8D^{4[AB][CD]}V_{[AB][CD]}.}
This relation is in fact telling us the usual story that the conformal supergravity gauge potential $V_{[AB][CD]}$ is {\it conjugate} to the {\it supercurrent} multiplet $J_{[AB][CD]}$. It has the same value for the Casimir since $E_0\rightarrow 4-E_0$.

We then see that the proposed coupling $\int d^4x \phi_0F^2_{\mu\nu}$ \refs{\pol,\witten} is now replaced by the $N=4$ manifest superconformal coupling \stelle,

\eqn\coup{\int d^4xd^{16}\theta V^{[AB][CD]}J_{[AB][CD]}.}

This is the manifest $N=4$ supergravity coupling of the $N=4$ conformal supergravity
field to the {\it conformal operator} on the boundary.
The dilaton coupling
\eqn\dil{\int d^4x \phi_0F_{\mu\nu}^2,}
considered in \refs{\pol,\witten} comes, for example, from the $\theta^{12}$ component of $V$
multiplied by the $\theta^4$ component of $W$ to match the $\theta^{16}$
integrand.

\lref\fggp{S. Ferrara, R. Gatto and  A. F. Grillo, Springer Tracts in Modern Physics, vol. 67 (Berlin-Heidelberg), New York, Springer 1973.}
\lref\fggptwo{S. Ferrara, R. Gatto, A. F. Grillo and G. Parisi, in {\it Scale and Conformal symmetry in hadron physics}, edited by R. Gatto, Wiley, New York, 1983.}
\lref\luscher{G. Mack and M. Luscher, Comm. Math. Phys. 41 (1975) 203.}
\lref\macktwo{G. Mack, J. Phys. 34, Colloque C-1 (Suppl. au no. 10) (1973) 79.}
\lref\mackthree{G. Mack and I. Todorov, Phys. Rev. D8 (1973) 1764.}

Note that $V^{[AB][CD]}$ has a very large negative conformal dimension ($-6$) but, due to its large gauge symmetry, in the Wess-zumino gauge, it starts with the $\theta^{12}$ component which is precisely $\phi_0$.
\lref\howe{P. S. Howe and  P. C. West, {\it Non-perturbative Green's functions in theories with extended superconformal
     symmetry}, hep-th/9509140; Phys.Lett. B389 (1996) 273, hep-th/9607060; {\it Is N=4 Yang-Mills Theory Soluble?}, hep-th/9611074; {\it Superconformal Invariants and Extended Supersymmetry}, hep-th/9611075.}
\lref\park{J. H. Park, {\it N=1 Superconformal Symmetry in Four Dimensions}, hep-th/9703191.}
 
There is, therefore, a manifest $N=4$ setting of the CFT/AdS proposal made in \refs{\pol,\witten}: if the
value at the boundary of the conformal supergravity field $V$, that couples to the conformal supercurrent $J$, is extended to a field $\hat V$ in the bulk of $AdS_5$, the CFT generating
functional with sources $V$ is identified with the supergravity action evaluated on $\hat V$ in the bulk:
\eqn\gen{Z(V_{[AB][CD]})_{CFT}=\langle e^{i\int d^4xd^{16}\theta V^{[AB][CD]}J_{[AB][CD]}}\rangle_{CFT}= Z_{AdS}(\hat V^{[AB][CD]}).}
$Z(V_{[AB][CD]})$ is the generating functional for the CFT correlators of superconformal currents. The previous formula is therefore a non-trivial statement
which allows in principle to compute CFT Green functions and OPE using tree-level supergravity theory. Conformal field theory correlators in $N=4$ superconformal Yang-Mills theory are highly constrained by OPE techiques. OPE in four-dimensional conformal invariant quantum field theories were widely studied in the seventies \refs{\fggp,\fggptwo,\macktwo,\luscher,\mackthree}  and, more recently, further generalized to
extended superconformal field theories including the $N=4$ case \refs{\howe,\park}. 
  
The $E_0=2$ scalar is the last component of $V^{[AB][CD]}$ which couples to the $E_0=2$ composite in $J_{[AB][CD]}$. This is obvious from superfield multiplication. In an analogous way, there is a $E_0=1$ scalar coupled to the $E_0=3$ scalar composite. 
\appendix{A}{Useful formulae for supercurrents and conformal supergravity fields.} 
In this appendix we summarize some results of refs. \refs{\roo,\stelle}
concerning supercurrents and their coupling to the conformal supergravity fields.

\lref\wess{J. Wess and B. Zumino, Nucl. Phys. B77 (1974) 73.}

For the sake of simplicity, we will consider the two cases of $N=1$ and $N=4$
conformal supergravity, the former because of its simplicity, being related to $N=1$ supergravity, and the latter because of its relevance in the present context.

\lref\sierra{M. Gunaydin, G. Sierra and P. Townsend, Nucl. Phys. B253 (1985) 573.}

Let us start with the $N=1$ case \fz. The superconformal algebra  in this case is $U(2,2/1)$ \refs{\fe,\wess} and the supercurrent is a vector real superfield $J_{\alpha\dot\alpha}$ satisfying the constraint $D^{\alpha}J_{\alpha\dot\alpha}=0$. It can be shown that this implies that the only field components of $J_{\alpha\dot\alpha}$
are $(A_\mu,J_{\mu\alpha},T_{\mu\nu})$, i.e. an axial current, a vector spin current and a symmetric tensor which are conserved conformal fields,
\eqn\currents{\eqalign{\partial^\mu A_\mu =0\qquad &A_\mu: D(3,{1\over 2},{1\over 2})\cr \partial^\mu J_{\mu\alpha}=\gamma^\mu J_{\mu\alpha}=0\qquad &J_{\mu\alpha}: D({7\over 2},1,{1\over 2})+D({7\over 2},{1\over 2},1)\cr
T_{\mu\mu}=\partial^\mu T_{\mu\nu}=0\qquad &T_{\mu\nu}: D(4,1,1)}}
The above is precisely the massless representation of supergravity on $AdS_5$ with an $O(2)$ gauge group, related to the $U(1)$ R-symmetry of the boundary theory \sierra.

The Weyl supermultiplet is a chiral $({3\over 2},0)$ representation of $SL(2,C)$ $W_{\alpha\beta\gamma}$ with conformal weight $3/2$. It can be expressed as $W_{\alpha\beta\gamma}=\bar D^2D_{\{\beta}\partial_{\alpha\dot\alpha}V^{\dot\alpha}_{\gamma\}}$, with a {\it prepotential} $V_{\alpha\dot\alpha}$ (of weight $-1$)
which corresponds to the gauge potential of $N=1$ conformal supergravity \fztwo. The
{\it current} coupling is therefore in this case,
\eqn\couplingcurr{\int d^4xd^4\theta V^{\alpha\dot\alpha}J_{\alpha\dot\alpha}}
In this case we take as  {\it singleton multiplet} the $N=1$ Yang-Mills multiplet $D(2,1,0|0)+D(3/2,1/2,0|3/2)+D(3/2,0,1/2|-3/2)$ (where the fourth index is
the $U(1)$ charge) and $J_{\alpha\dot\alpha}$ can be written as a composite field as 
\eqn\comp{J_{\alpha\dot\alpha} = Tr W_\alpha W_{\dot\alpha},}
where $W_\alpha$ is the chiral field strength of the singleton (Yang-Mills)
multiplet\foot{In this case, one can have additional singletons, the chiral multiplets, with $D(E_0,0,0|E_0)+D(E_0+1/2,1/2,0|E_0-3/2)$.}.

In the $N=4$ case, the singleton superfield is a $N=4$ superfield $W_{[AB]}$ which satisfies the constraints given in \stelle\foot{Note, in particular, that, unlike the $N=1,2$ cases, $W_{[AB]}$ is not chiral.}. In terms of $W_{[AB]}$ ($A=1,...,4$) which can
also be written (as we did in section 4) as $W_l, (l=1,...,6)$ (vector of
$SU(4)=O(6)$) the supercurrent multiplet $J$ is a composite superfield given by,
\eqn\dual{J_{mn}= Tr (W_lW_n - {1\over 6}\delta_{mn}W^pW_p),}
i.e. in the ${\bf 20}_R$ of $SU(4)$.

The Weyl multiplet is a chiral superfield which admits a {\it prepotential}
$V_{lm}$ such that 
\eqn\curv{W=\bar D^8D^{4lm}V_{lm},}
where $V_{lm}$ has conformal weight $-6$. The current coupling in the $N=4$
case is then \stelle
\eqn\cc{\int d^4xd^8\theta d^8\bar\theta V^{lm}J_{lm}.}
For completeness we give here the explicit expression of $J_{lm}=J_{ABCD}$
as a composite of singleton superfields. This was obtained in \roo. The components of the multiplet are $(g_{\mu\nu}, \psi_{\mu}^A, A_{\mu B}^A, A_{\mu\nu}^{AB},\lambda_A,\chi^{AB}_C, \phi, e_{AB}, d^{AB}_{CD})$ i.e. a symmetric tensor current, a vector spin current in the ${\bf 4}+\bar{\bf 4}$, a vector current in the adjoint of $SU(4)$, an antisymmetric in the ${\bf 6}_c$, fermions $\lambda$ and $\chi$ in the ${\bf 4}$ and ${\bf 20}$, scalars $\phi$,$e$ and $d$ in the ${\bf 1}$, ${\bf 10}$ and ${\bf 20}_R$. 
In terms of the $N=4$ supermultiplet components, $(F_{\mu\nu},\psi_A,\phi_{AB})$ (where $\psi_A$
are the four spinors rotated by $SU(4)$ and $\phi_{AB}$, subject to the constraint $\phi^{AB}=(\phi_{AB})^*={1\over 2}\epsilon^{ABCD}\phi_{CD}$, are the scalars in the ${\bf 6}$ of $SU(4)$),   the explicit expression is (formulae (A.4) and
(A.6) in \roo )
\eqn\explicit{\eqalign{g_{\mu\nu}=&{1\over 2}\{\delta_{\mu\nu}(F^-_{\rho\sigma})^2 -4F^-_{\mu\rho}F^-_{\nu\rho} + h.c.\} - {1\over 2}\bar\psi^A\gamma_{\{\mu}\partial^{\leftrightarrow}_{\nu\}}\psi_A +\delta_{\mu\nu}(\partial_\rho \phi^{AB})(\partial_\rho \phi_{AB})\cr
&-2(\partial_\mu \phi^{AB})(\partial_\nu \phi_{AB})-{1\over 3}(\delta_{\mu\nu}\partial^2 -\partial_\mu\partial_\nu )(\phi^{AB}\phi_{AB}), \cr
\psi_{\mu}^A=&-(\sigma F^- )\gamma_{\mu}\bar\psi^A + 2i\phi^{AB}\partial^{\leftrightarrow}_\mu\psi_B + {4\over 3}i\sigma_{\mu\lambda}\partial_\lambda (\phi^{AB}\psi_B ),\cr
A_{\mu B}^A =&
\phi^{AC}\partial^{\leftrightarrow}_{\mu}\phi_{CB} + \bar\psi^A\gamma_\mu\psi_B -{1\over 4}\delta^A_B \bar\psi^C\gamma_\mu\psi_C,\cr
A_{\mu\nu}^{AB}=&\bar\psi^A\sigma_{\mu\nu}\bar\psi^B + 2i\phi^{AB}F^+_{\mu\nu},\cr
\lambda_A=&\sigma F^- \psi_A,\cr
\chi^{AB}_C=& {1\over 2}\epsilon^{ABDE}(\phi_{DE}\psi_C +\phi_{CE}\psi_D ),\cr
 \phi =& (F^-_{\mu\nu})^2,\cr
 e_{AB}=&\psi_A\psi_B,\cr  d^{AB}_{CD}=& \phi^{AB}\phi_{CD}-{1\over 12}\delta^{[A}_C\delta^{B]}_D\phi^{EF}\phi_{EF}.}}
The currents are conserved and satisfies,
\eqn\cons{\eqalign{ \partial_\mu g_{\mu\nu}&=0\qquad\qquad \partial_\mu \psi_{\mu}^A=0,\cr g_{\mu\nu}&=g_{\nu\mu}\qquad\qquad \gamma_\mu \psi_{\mu}^A=0,\cr
g_{\mu\mu}&=0\qquad\qquad \partial_\mu A_{\mu B}^A=0.}}

\centerline{\bf Acknowledgements}
S. F. is supported in part by DOE under grant DE-FG03-91ER40662, Task C, and by ECC Science
Program SCI*-CI92-0789 (INFN-Frascati)

\listrefs
\end